\documentclass[a4paper,12pt]{article}
\usepackage[british]{babel}             % British English hyphenation
\usepackage[margin=3cm]{geometry}       % margins
\usepackage{newtxtext}                  % Good fonts
\usepackage[slantedGreek]{newtxmath}    %   "    "   (slanted Greek)
\usepackage[hyphens]{url}               % for URLs (allow linebreak at hyphens)
\usepackage{hyperref}
\DeclareUrlCommand\doi{\urlstyle{rm}}   % \doi command
\renewcommand{\doi}[1]{\href{https://doi.org/#1}{#1}}
\usepackage{natbib}                     % normal bibliography
\usepackage{RSBMbib}                    % for simple bibliography

\usepackage{tikz}            % needed for floating boxes (calls in graphicx.sty)
\usepackage{graphicx,color}

	\newlength{\figwidth}
	\newlength{\shift}
	\newcommand{\fg}[3]
	{
		\begin{figure}[ht]
			
			\vspace*{-0cm}
			\[
			\includegraphics[width=\figwidth]{#1}
			\]
			\vskip -0.2cm

			\caption{\label{#2}
				\small#3
			}
	\end{figure}}
	
\title{Philip ~Warren ~Anderson\\
     13 December, 1923 - 20 March 2020\\ 
       Elected FRS 1980}

\author{Premala Chandra $^{1}$, Piers Coleman$^{1,2}$ and Clare C. Yu$^{3}$.\\
\\
$^1$\emph{\small Center for Materials Theory, Department of Physics and Astronomy,}\\[3pt]
\emph{\small Rutgers University, 136 Frelinghuysen Rd., Piscataway, NJ 08854-8019, USA.}\\[6pt]
$^2$\emph{\small Department of Physics, Royal Holloway, University of London, Egham, Surrey TW20 0EX, UK.}\\[6pt]
$^3$\emph{\small Department of Physics and Astronomy, University of California, Irvine, Irvine, CA 92697, USA}\\[6pt]}

\begin{document}

\vspace*{-0cm}
			\[
			\includegraphics[width=1.1\textwidth]{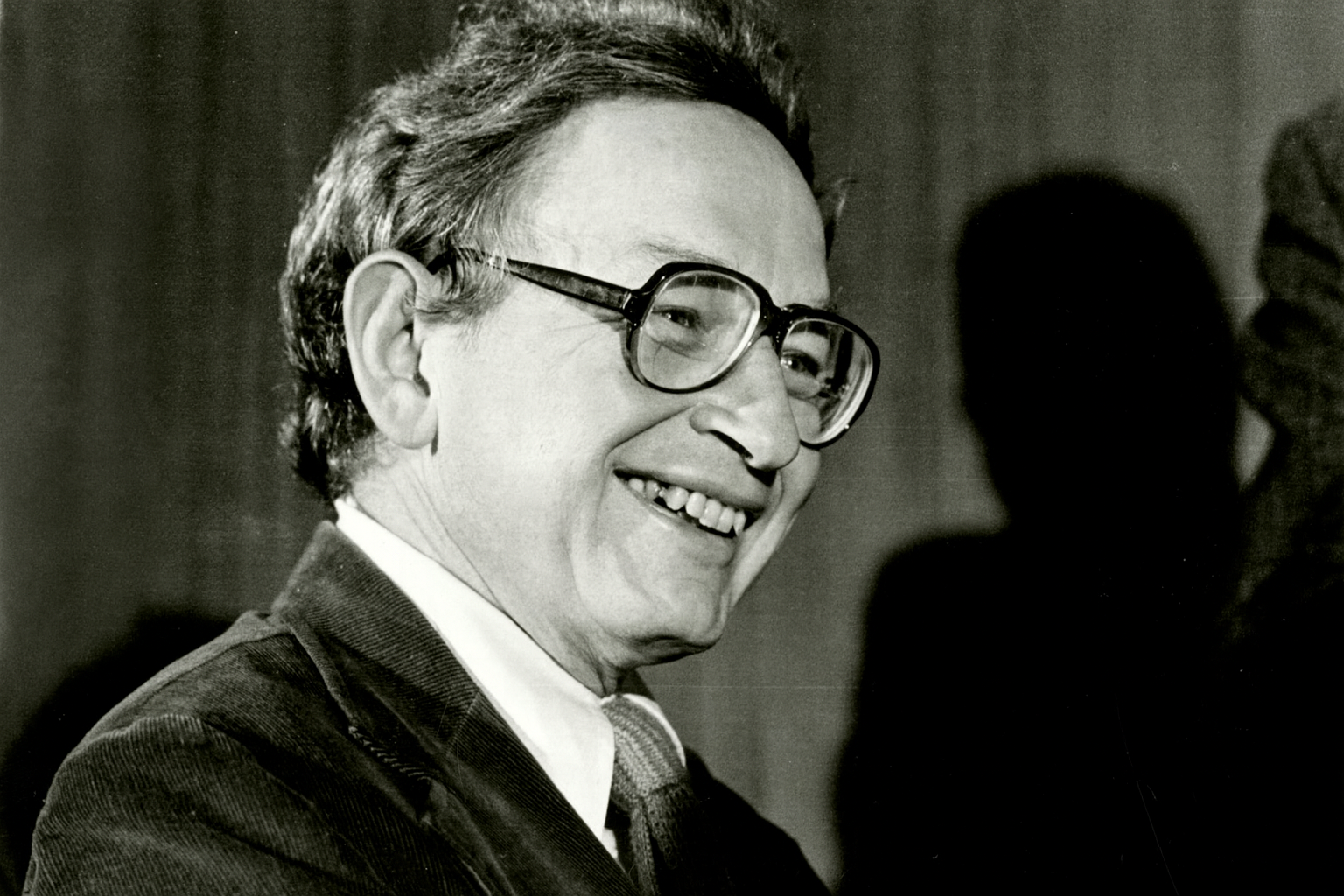}
			\]
			\vskip -0.2cm

%			\caption{\label{}
				%\small#Philip Warren Anderson
%			}

\vskip0.3truein

\begin{center}
{\bf Philip Warren Anderson}   
\end{center}

\maketitle

%%%% Subject entries to be placed here %%%%
%\subject{biographical memoir,condensed matter physics}

%%%% Keyword entries to be placed here %%%%
%\keywords{Anderson Higgs, Emergence, Localization, Disorder, Superconductivity}

%%%% Insert corresponding author and its email address}
%\corres{Insert corresponding author name\\
%\email{cyu@uci.edu}}

%%%% Abstract text to be placed here %%%%%%%%%%%%
\begin{abstract}
Philip Warren Anderson was a pioneering theoretical physicist whose work fundamentally shaped our understanding of complex systems. Anderson received the Nobel Prize in Physics in 1977 for his groundbreaking research on localization and magnetism, yet he did so much more. His work on magnetism included antiferromagnetism, superexchange, the Kondo problem and local magnetic moments in metals. Anderson pointed out the importance of disorder through his work on localization, non-crystalline solids and spin glasses. In superconductivity, he is known for the dirty superconductor theorem, showing the gauge-invariance of the  BCS theory, 
%for explaining how repulsive electrons could form Cooper pairs, 
his study of flux creep,
and for his collaboration with experimentalists to realize the Josephson effect.
Anderson's resonating valence bond theory may yet play an important role in high temperature superconductivity. 
%With regards to superfluidity, the A-phase of superfluid helium-3 was foreseen by his model with Morel. 
Anderson was also fascinated by broken symmetry, 
and he laid the theoretical groundwork for what is now known as the Anderson-Higgs mechanism, showing how gauge bosons can acquire mass —an insight that played a foundational role in the Standard Model of particle physics. In his seminal ``More is Different'' paper, Anderson argued that the collective emergent phenomena that arise in complex interacting systems cannot be deduced from their fundamental parts.  Anderson’s legacy endures not only through the lasting impact of his scientific work but also through his influence on generations of physicists who continue to explore the rich landscape of collective behavior in nature.
\end{abstract}
%%%%%%%%%%%%%%%%%%%%%%%%%%%
\pagebreak

% Do not delete this null reference, as it puts the Anderson references into the biography 
%\nocitePhil{5Anderson,50AndersonSuperexchange,RN3430,48Anderson,66Anderson,99AndersonRPA,36Anderson,75Anderson,37Anderson,90AndersonHe3,76Morel,78Anderson,67Anderson,68Anderson,notebook,13Anderson,40Anderson,70AndersonPoorManScaling,23Anderson,61Anderson,44Anderson,80Fazekas,117Edwards,77AndersonTAP,2Anderson,16Anderson,19Abrahams,49Fleishman,9Anderson,20Anderson,82AndersonRVB,28Anderson,95AndersonNeutron,21Anderson,83AndersonRVBvanilla,65Anderson,6Anderson,1031Anderson,22Anderson,74Anderson,98AndersonHiggs}}
\part{Biographical Sketch}

\section{{\bf Early Years 1923 -- 1949}} Philip Warren Anderson (``Phil") was
born on December 13, 1923, in Indianapolis, Indiana; he grew up in
Urbana, Illinois where his father, Harry W. Anderson, was a
professor of plant pathology at the University of Illinois (UIUC). His paternal
grandfather was a fire and brimstone preacher who later turned to
farming. Phil's maternal grandfather was a professor of mathematics at
Wabash College (Indiana) where his maternal uncle, a Rhodes Scholar, was
later on the faculty \citep{1Bernstein}. Phil's mother, Elsie O. Anderson (``Bodie"), was a formidable woman who had great plans for Phil. In later years Phil said that his older sister Eleanor Grace   Maass (``Graccie") should have received  similar attention as she too was very bright. Phil's daughter, Susan, notes that when Bodie learned %that 
Harvard required two years of Latin \citep{1Bernstein}, ``she exerted her considerable magnetism" to have the chairman of the UIUC Classics Department tutor Phil \citep{1Bernstein}, and that without her ``determined support and encouragement, he would have never made it to Harvard!" Phil's parents belonged to a warm group of
friends known as the ``Saturday Hikers''; some of his fondest childhood
memories were of their outdoor 
activities together (\ref{Anderson16}). The group was quite
politically conscious with a sense of foreboding about global pre-World
War II events. As a result, Phil retained strong political convictions
throughout his life. In 1937, during his father's  sabbatical, Phil
spent an impressionable year in Europe, and much later he often spoke
warmly about it.

In Urbana, Phil attended University High School where a math teacher,
Miles Hartley, was one of the few who challenged Phil intellectually.
Subsequently Phil headed to Harvard in 1940 on a National Scholarship,
intending to major in mathematics. In a letter responding to Harvard's
request for a profile of his son, Harry Anderson describes Phil as an
``{\sl even-tempered boy, tolerant of others' opinions but likely to defend
his own stubbornly\ldots He is not at ease with people who do not
interest him}'' \citep{3Zangwill}. The subsequent entry
of the US into World War II shortened Phil's college years; he graduated
with a degree in ``Electronic Physics'' in 1943 and built
antennas at the Naval Research Laboratory (NRL). This experience left
him with a lasting admiration for Bell
Labs engineers. While at the NRL, Phil received a quantum mechanics
textbook from a colleague as collateral for money owed, and he eventually
kept it.

After the War Phil returned to Harvard to pursue a PhD in physics. In
addition to his doctoral research, Phil enjoyed doing puzzles and singing with friends including Tom Lehrer, who
later became famous for his sardonic songs. While home visiting his
parents in Urbana, Phil met Joyce Gothwaite. They married shortly afterwards and a daughter, Susan, soon
followed. As Ravin Bhatt, a colleague at Bell and then Princeton, has 
noted \citep{Bhatt21}, ``{\sl Joyce was truly Phil's 'rock' who kept him grounded, taking care of everything so that Phil could do his physics without worries.}"

Inspired by John Van Vleck's lectures on solid state physics, Phil
decided to do his dissertation with him.
Phil's thesis
problem involved the observed line broadening in radio frequency spectra
that could be studied due to wartime advances in electronics. Phil
found that the techniques he was learning in  
quantum field theory were useful for his thesis research.
His theoretical results were in excellent agreement with experiment.
Phil successfully defended his thesis, ``The Theory of Pressure
Broadening of Spectral Lines in the Microwave and Infrared Regions,'' on
January 19, 1949, and later published his results in 
Physical Review (\ref{5Anderson}); this paper continues to be cited.

\section{{\bf Bell Labs 1949 -- 1959}} 
Van Vleck (``Van'') taught Phil to have a deep respect for experiment, to identify
key underlying concepts and to develop minimalist descriptive models
\citep{3Zangwill}; many years later Phil passed along this philosophy to his many
mentees under the mantra ``follow the data''. While pursuing his
doctoral work, Phil learned of several experiments performed at Bell
Labs, and he was determined to go there 
as a researcher (\ref{6Anderson}); thanks to
Van's efforts, in 1949 this happened. Whenever he reminisced about his
times at Bell Labs, Phil had a twinkle in his eye. This research branch was originally intended to support its
parent company, AT\&T, to create a global communications system.
However, it also developed into a center of innovative thinking, particularly in the then nascent
field of solid-state physics. ``{\sl At first sight,}'' noted the writer
Arthur C. Clarke in the 1950's, ``{\sl when one comes upon it in its
surprisingly rural setting \ldots [it] looks like a large and up-to-date factory, which in a sense
it is. But it is a factory of ideas.}'' \citep{8Clarke,7Gertner} The scientific environment at Bell Labs had an
incredible ambiance: it was a world of research unencumbered by
bureaucracy that is seldom experienced today. Collaborations between
researchers were encouraged, and chance interactions frequently occurred
in the long corridors and in the stairwells. Phil adapted quickly to
these surroundings, and many of his discoveries were connected to
his close interaction with experimentalists.

When Phil arrived at Bell Labs in 1949, quantum mechanics was not yet
twenty-five years old. In the
mid-1930s Eugene Wigner at nearby Princeton, with his students John
Bardeen, Conyers Herring and Fred Seitz, was establishing the field of solid-state physics. Shortly after World War II, Bardeen and Herring came to Bell Labs and
Seitz went to Illinois. Both Illinois and Bell became powerhouses of
solid-state physics in the USA. At Bell, Phil's mentors were Gregory
Wannier, Conyers Herring and Charles Kittel, theorists roughly ten years
older than him, who introduced him to magnetism, phase transitions and
current research developments elsewhere. In early 1956 the theorists
formed their own department with postdoctoral and summer visitor
programs. It was an extraordinary center of intellectual activity, and
Phil quickly became one of its central figures. He had a very
interactive style, constantly sharing his ideas as they were developing.
Morning coffees and regular group lunches were always very animated
affairs with lots of lively, sometimes heated, discussions. For young
aspiring researchers, these were wonderful opportunities to see physics
evolving in real time!

At Bell Phil transformed rapidly from a novice to a senior figure in
condensed matter physics. Later he always said that his early years
studying magnetism gave him the grounding and the flexibility to move to
other areas (\ref{9Anderson}). This was a period of tremendous research productivity
for Phil; here we provide just a flavor of his work with more discussion
to follow. Phil's detailed studies of antiferromagnetism led him to
broken symmetry, a deep concept in physics with relevance well beyond
the materials that inspired its original study. His interactions with
experimentalist Bernd Matthias led him to explain how electrons form
magnetic moments in alloys \ref{37Anderson}, part of the work later cited by the Nobel
Prize committee. When Bell Labs expanded into low temperature research,
Phil played key roles in the areas of superconductivity and
superfluidity. Furthermore, his study of broken symmetry in the presence
of long-range forces resulted in the Anderson-Higgs mechanism in
particle physics. Inspired  by George Feher's experiments on doped
semiconductors, Phil found that disorder in
materials could lead to new concepts rather than just being an
inconvenient nuisance. For example, Phil realized that disorder in
metals could ``localize'' electron waves \ref{48Anderson}, transforming them into
insulators, a process now called ``Anderson localization''; this work played a central role in his Nobel citation. Measurements
indicating that there were low-temperature random frozen spin states,
rather than conventional spin ordering, in a class of magnetic materials
now known as spin glasses, led Phil to appreciate that equilibrium
statistical methods were insufficient to describe these systems \ref{117Edwards}. The
approaches that his studies launched led to methods in combinatorial
optimization that are used in very large-scale integrated (VLSI) chip
design.

In the 1950s, the Bell Labs management realized that their scientists
would benefit from research sabbaticals away from the Labs. Phil took full advantage of this freedom, and he would later claim that it
was through his trips abroad that he was first
appreciated within the academic research 
community (\ref{1031Anderson}).
In 1953-54 Phil was invited by Ryogo Kubo at the University of Tokyo for
a sabbatical stay (\ref{6Anderson}). The Anderson
family sailed across the Pacific for a six-month visit. Japan was still
recovering from WWII, and large parts of Tokyo had yet to be rebuilt.
Anderson attended the 1953 International
Conference in Theoretical Physics in Tokyo and Kyoto, his first
experience on the international stage; here he made several
acquaintances that would influence his later career, including Nevill
Mott. Anderson and Kubo had adjoining offices on the Hongo campus of
the University of Tokyo, and regularly discussed their ideas on magnetism,
electron spin resonance and statistical mechanics.  
During this stay, Phil also 
learned to play the Japanese game of Go, which 
became a lifetime pastime; he eventually attained the rank of first-dan master and in
2007 he received a lifetime achievement award from the Nihon Ki-in, Japan's Go
association \citep{11Brinkman}. Later in life, Phil and Kubo would each jokingly claim that they were responsible for discovering
the other one. Phil would also proudly recall pulling Kubo out of his office
after he had become unconscious due to a gas leak.
\figwidth=0.95\textwidth
\fg{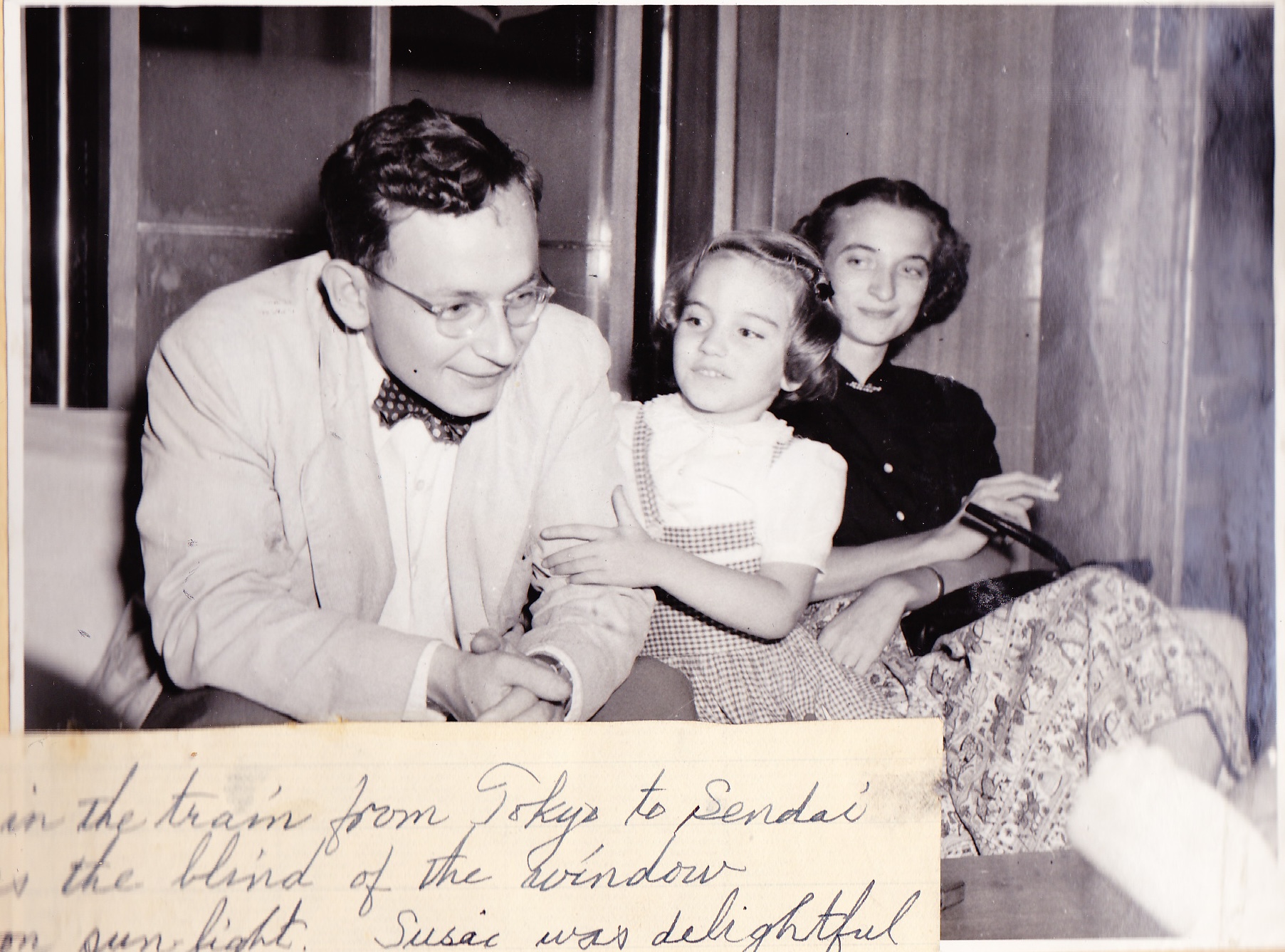}{fig1x}{Phil and Joyce Anderson with their daughter Susan in Japan, 1951. (Courtesy, Susan Anderson) }

\section{{\bf Cambridge and England 1959 -- 1975}} In 1959 Phil
was invited to a conference on magnetism held at Brasenose College,
Oxford. There Phil learned about Jacques Friedel's and Andrei Blandin's ideas on magnetic moment formation in metals, an approach that Phil would later
transform into Anderson's local moment 
model \ref{37Anderson}. During this trip, interest in
bringing Phil to England for an extended stay 
grew \citep{12Anderson}. In
1961-62, Phil was invited by Brian Pippard to spend a year in Cambridge. Anderson lectured on solid
state physics, many body theory and superconductivity in a course
attended by a beginning graduate student, Brian Josephson. After one
such class, Josephson showed Phil his calculations of Cooper pair
tunneling, work that heralded the discovery of the Josephson effect for
which Josephson would be awarded a Nobel Prize in 1973. It is a testimony to
Phil's character that he always gave Josephson full credit for the
discovery (\ref{13Anderson}). Brian Josephson warmly remembers the many
discussions they had together, noting
``{\sl one amusing thing was he couldn't be
bothered to get equations exactly right, so had a convention whereby $\pi$
and i were considered equal to one so they could be omitted from the
equations!}\ ''

Phil's sabbatical year went sufficiently well that Cambridge colleague Volker Heine recalls 
``{\sl I can say that Mott and Pippard were keen to have Phil here half-time, shared with Bell Labs, more than anyone else full-time}," and thus
Phil became a part-time Professor there from 1967 to 1975.
This was an extraordinarily productive time for Phil.  
Concurrently Phil developed his philosophy of
``More is Different'' where he advocated that science is hierarchical,
with new emergent principles to be found at each level of complexity. T.V. Ramakrishnan, a longtime friend and collaborator writes \citep{Rama20} that ``{\sl It awoke us to the reality
of emergence, at a stage when 'real science' was equated
with reductionism.}"
During his tenure at Cambridge, Phil mentored several doctoral
students, visitors and postdocs
including Duncan Haldane, Michael Cross, John Inkson, Richard Palmer,
David Bullet, Gideon Yuval, Ali Alpar and John Armitage,
Erio Tosatti, Dennis Newns and Patrick Fazekas. Erio recalls
that ``{\sl When Phil returned from the US it felt like a ray of
sun\ldots Interactions with Phil were, in everybody's opinion, a special experience -- treasured by some, feared by others. Despite his kindness and human concern for everyone young and in any way helpless, Phil was
generally a cryptic communicator, for some a mumbler, not easy to understand. If you, like myself, belonged to the first type of people, it was enough to listen to him for a quarter of an hour, to become
instilled with so many ideas and insights to keep you busy for weeks and
months.}''

Phil's stay at Cambridge led to major progress across a diverse range of
topics from neutron stars to spin glasses, while he also wrote the text 
``Concepts in Solids'' (\ref{anderson1972concepts}) based on his graduate lectures.  With Yuval, Phil
developed a new renormalization approach for the physics of magnetic ions
in metals known as the ``Kondo problem''. He worked on neutron stars and charge density waves with Alpar, Palmer and Tosatti.
With Haldane, Phil developed early theories of valence fluctuations in
solids; he worked on superfluid Helium-3 with Cross; and with Fazekas,
Phil developed the concept of spin liquids which would later become the
basis of his theory of high temperature superconductivity. A particularly notable
collaboration developed with his colleague Sam Edwards on Saturday
mornings. At that time Edwards was Chairman of the Science Research Council, a 
UK funding body, and only returned to Cambridge on weekends. Phil and
Edwards met every Saturday morning,
developing a theory of spin freezing in random magnets called spin
glasses. Concurrently Phil and his Cambridge
colleague Volker Heine renamed
their former Solid State theory group as that of Condensed Matter; this nomenclature is now used everywhere.

Many researchers had some of their most memorable interactions with Phil
outside the Cavendish lab. Volker and Erio recall that Phil would
always eat a pub lunch with colleagues, students and postdocs, providing an opportunity for regular scientific and cultural interactions.
Phil and Joyce Anderson were also wonderful hosts, and many fondly remember the parties they held in
Cambridge during this period. In 1973 Phil and Joyce bought a holiday cottage in Port Isaac, Cornwall; postdocs and visitors were often invited to visit them there. Physics
conferences were also venues for being with Phil in a more relaxed
fashion. One such memorable occasion occurred at the 1974 August Summer
School for Low Temperature Physics where the recent discovery of
superfluidity in Helium-3 was discussed. The workshop took place in St
Andrews in Scotland. %Erio recalls \citep{15Tosatti}``He-3 was the highlight at that school. \ldots With Phil, Ann [Eggington],
%Mike Cross, and others we went off hiking in rocky hills and
%peat-covered valleys, bathing in the nude (except Phil) in ice-cold
%brooks. We also indulged in single-malt whisky outings --- %most notably
%one in~Auchtermuchty where we got considerably merry %(including Phil)
%with running contests among us''%
There Ann Eggington, conference attendee,  recalled an excursion with Phil
 (see Fig. (\ref{fig2}):``{\sl We all set off to a boat trip~to Puffin Island. The weather was
agitated and the captain would not unmoor, seeing us as a human cargo of
dubious resilience to the rough sea. As it happened, the captain's name
was Anderson. For Erio and I it was an easy game to ask Phil Anderson
to confront Captain Anderson and guarantee that we would be OK ~--- and
it worked!}\ ''

\figwidth =0.95\textwidth
\fg{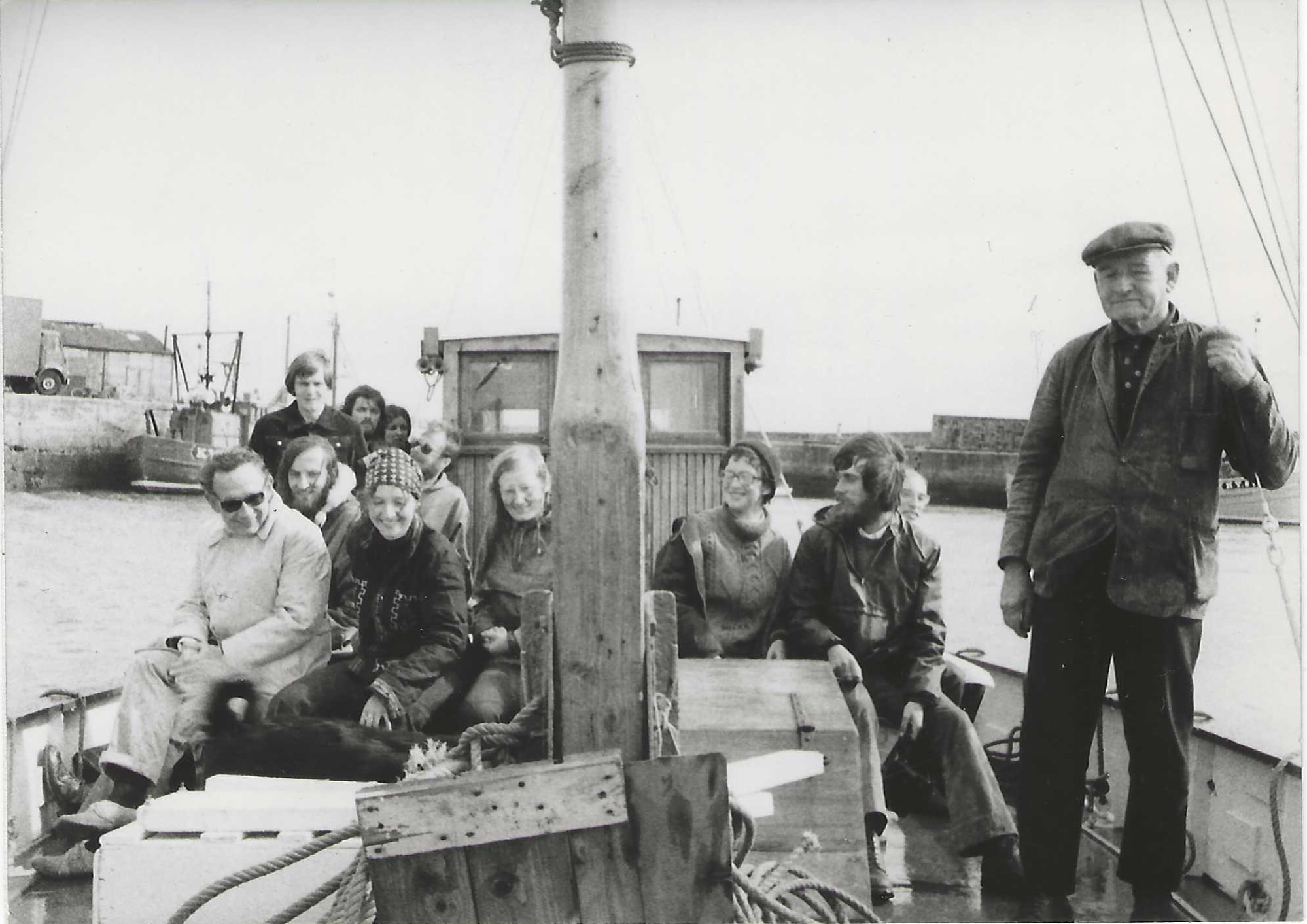}{fig2}{Phil Anderson (left) with students Richard Palmer(behind Phil), Ann Eggington (right of Phil) and others on an  excursion to Puffin island as part of the St Andrew's 1974 School on Low Temperature Physics. (Photo courtesy of Ann Eggington). }

\section{{\bf Princeton 1975 Onward} }After eight years of feeling like tourists
in both the British and American cultures, ``with no really satisfactory
role in either,'\,' Phil and Joyce returned to the United States (\ref{Anderson16}). Phil traded his part-time appointment at Cambridge for a part-time
position as the Joseph Henry Professor of Physics at Princeton
University in the fall of 1975, becoming full time at Princeton from
1984 to 1997 when he became Professor Emeritus. His new position allowed
him to continue his work at Bell Labs, while establishing a center for
condensed matter at Princeton University. Phil's arrival there quickly
nucleated a buzz of research activity: new students eagerly lined up to
work with him, including Carol Morgan Pond, Daniel Stein, Henry
Greenside and Albert Chang, joining Richard Palmer, Ali Alpar and Duncan
Haldane who traveled with him from Cambridge.
James Sethna, Khandker Muttalib, Clare Yu, Gabriel Kotliar,
Piers Coleman, Yaotian Fu, Al Kriman, and Jerry Tesauro were among a subsequent group of students; a later set
included Ted Hsu, Joe Wheatley, Zou Zhou, Jonathan Yedida, Philip Casey,
Charles Stafford, Steve Strong, and visitors Ganapathy Baskaran, Premi Chandra, Benoit
Doucot, Antoine Georges, Ido Kanter and Shoudan Liang. There was much to learn from Phil.  Ganapathy Baskaran notes that ``{\sl On several occasions, Anderson brought out deep meanings in ordinary sounding seminar talks and pleasantly surprised the speaker and the audience.}"

Phil cared deeply about the young researchers in his charge. 
However, communicating with Phil
was difficult which made working with him a challenge. 
``{\sl Phil was known...as a Delphic Oracle, given his tendency to make cryptic and puzzling remarks when discussing a physics problem....they often seemed mysterious at the time or else totally off the subject,}" notes Daniel Stein. 
``{\sl But then, weeks later, it would suddenly dawn on me what he meant...Once understood it always turned out to be insightful, relevant and deep.}" James Sethna recalls that ``{\sl posed with a mystery or a challenge, Phil would invent the
most interesting and insightful possible explanation.}"
He certainly had an unconventional approach.
Clare Yu remembers a time when Phil was
particularly exasperated with her and said, 
``{\sl Theoretical physics isn't [about] doing calculations. It's setting up the
problem so that any fool could do the calculation.}'' Gabriel
Kotliar shares that "{\sl Phil was an unusual PhD advisor...Pursuing hard problems without guaranteed success, together with  trusting with some skepticism intuitive reasoning  and creative methodologies  was Phil’s real lesson at the end, but\ldots it took some time to sink in.}"  

In October 1977 an early morning phone call from Stockholm informed
Phil, reputedly brought in from his garden, that he had won the Nobel
Prize for his work on electrons in disordered materials and his local
moment theory of magnetism. Phil shared the Prize with his doctoral advisor, John
Van Vleck, and with Nevill Mott, his old friend from 
Cambridge  (\ref{Anderson16}). The
acclaim, honor and attention of a Nobel Prize distracts many a researcher, but within a month of the festivities in Stockholm, Phil was
back to commuting between his fledgling group at Princeton and his
colleagues at Bell Labs. Indeed in the following years Phil and three
collaborators resolved several issues related to Phil's Prize-winning
work on electron localization leading to a publication 
in 1979 (\ref{19Abrahams}); the
team became known as the ``Gang of Four'', a reference to four political defendants in a trial in China at the time. The paper ignited renewed interest in the
field, and a slew of theoretical and experimental papers followed.

Phil's deep curiousity and passion for physics was undiminished by his
many laurels and subsequent commitments. Here we give a flavor of his
later adventures. Motivated by several experimental developments, Phil
continued to work on various forms of magnetism in the 1980's; he also
applied key concepts of spin glasses to combinatorial optimization and
NP completeness. In 1983, Phil published a graduate textbook, Basic
Notions of Condensed Matter Physics (\ref{20Anderson}), combining his unique
perspective of key ideas in the field with a selection of important
reprints. The discovery of high-temperature superconductivity in the
late 1980's remained a focus of Phil's attention for the remainder of his
life. Ten years later, Phil published another book, A Career in
Theoretical Physics, consisting of reprints of his selected articles
that are not easily accessible (\ref{21Anderson}). In 2011, Phil published his last
book, More and Different: Notes from a Thoughtful 
Curmudgeon (\ref{22Anderson}); the
title is a play on that of his famous 1972 Science article, ``More
is Different'' (\ref{23Anderson}) which refuted the reductionist approach to science
by noting that unexpected emergent phenomena can come from interactions
between simple objects. The book is a collection of his non-technical
writings including essays, personal recollections and book reviews,
giving a window into his thoughts and opinions about a broad
range of topics. Phil's daughter Susan notes
that her father had a tremendous sense of fun and that ``{\sl his love of the odd and the eccentric led to so many of his diverse interests.}'' She recalls that ``\ {\sl `higgledy-piggledy' was 
a favorite descriptor.}''

\fg{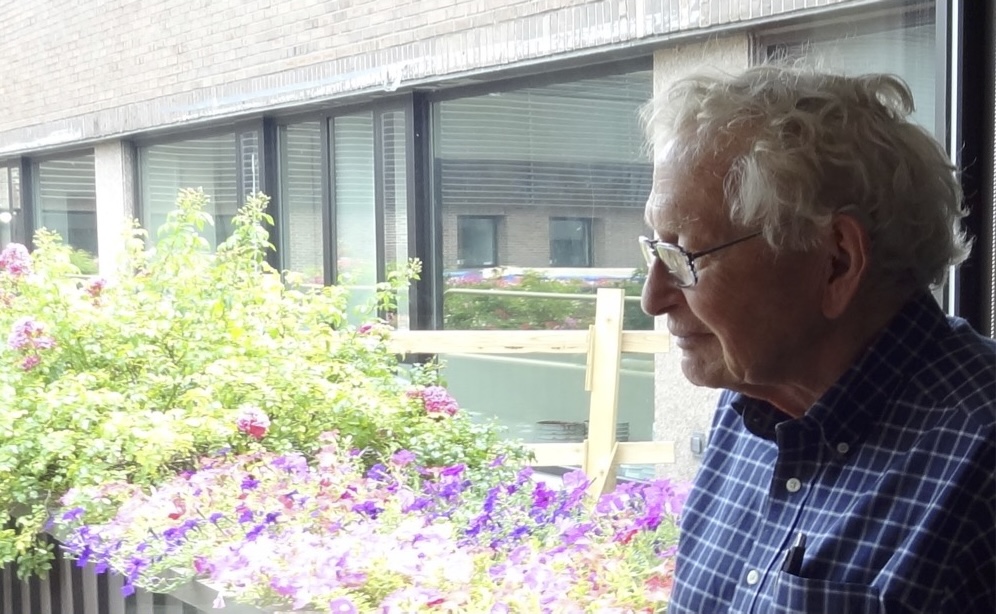}{figxx}{Phil Anderson at Jadwin Hall, Princeton 2011. (Photo courtesy Ganapathy Baskaran). }
In 1984 Phil retired from Bell Labs. After he became an emeritus
professor at Princeton in 1996,  he continued to be an  inspiring
presence in the department. ``{\sl Being Phil's colleague for the past
quarter century --- when he was already a legend --- was an honor in
itself,}'' says Shivaji Sondhi, who was a faculty colleague at Princeton \citep{13Zandonella} ``{\sl My
experience of him was of a man of~wide learning well beyond physics,
broad interests across science, enormous creativity and an extraordinary
capacity --- almost literally to the very end ---~to get up and think
about important problems in physics}''. In 2013 Phil's
90\textsuperscript{th} birthday was celebrated with a weekend workshop
that resulted in a book \emph{PWA90: A Lifetime of Emergence} \citep{Chandra15}; Phil
sat in the front row and asked questions of almost every speaker.
``{\sl There are very few people in condensed matter physics who have not
been influenced by Phil's ideas,}'' states Anthony Leggett (FRS 1980), a professor of
physics at the University of Illinois at Urbana-Champaign who received
the 2003 Nobel Prize in physics \citep{13Zandonella} ``{\sl Even when he turned out to be wrong,
he was influential because he got people to think in new directions.}''

Over the course of his career, Phil was a mentor and an inspiration to
many young researchers who themselves went on to make serious
scientific contributions. In 2016 one of Phil's former students and later a
Princeton professor, Duncan Haldane, was awarded the Nobel Prize in
Physics. Commenting on his doctoral days with Phil, Duncan notes \citep{13Zandonella} ``{\sl I had
the great fortune to have had him as my mentor when I was a graduate
student.~~I would regularly meet him to talk about the problem he had
given me to work on, but instead he would tell me about the things he
was thinking about that day, and seeing his thought process was an
amazing lesson in how to think about problems that decisively shaped my
future career.~~What a mentor!}'' 

\section{{\bf Bell Labs, Aspen and Santa Fe}} The demise of Bell Labs was
 heartbreaking for Phil: he took it personally as he had played a key role in building its international reputation. Opinionated as always, Phil often said
that Bell was ``{\sl extraordinarily poor at economically exploiting its
technology}'', and he felt that ``{\sl if managed through the post-84' crisis
with the flexibility and intelligence exhibited in those early
days\ldots.it might still be with us.}'' \ref{6Anderson}. 
Phil played important parts in two other research centers, 
the Aspen Center for Physics and the
Santa Fe Institute, both located in the
American West.

The Aspen Center for Physics (ACP), originally a satellite of the Aspen
Institute, became an independent entity in 1962.
Initially founded by high-energy physicists, Phil played a major role in
broadening its scope to include condensed matter workshops. 
Phil first
came to the ACP in the summer of 1975, and he would subsequently return
for a total of 22 summers, even serving as Chair of its Board of
Trustees (1982-1986). It was at the ACP that Phil worked with David
Pines, Gordon Baym and others to apply concepts of condensed matter
theory to astrophysical phenomena in neutron stars. In 2000 the ACP
held a winter workshop honoring Phil's scientific contribution; it
resulted a collection, \emph{More is Different: Fifty Years of Condensed
Matter Physics} \citep{27More}.

\fg{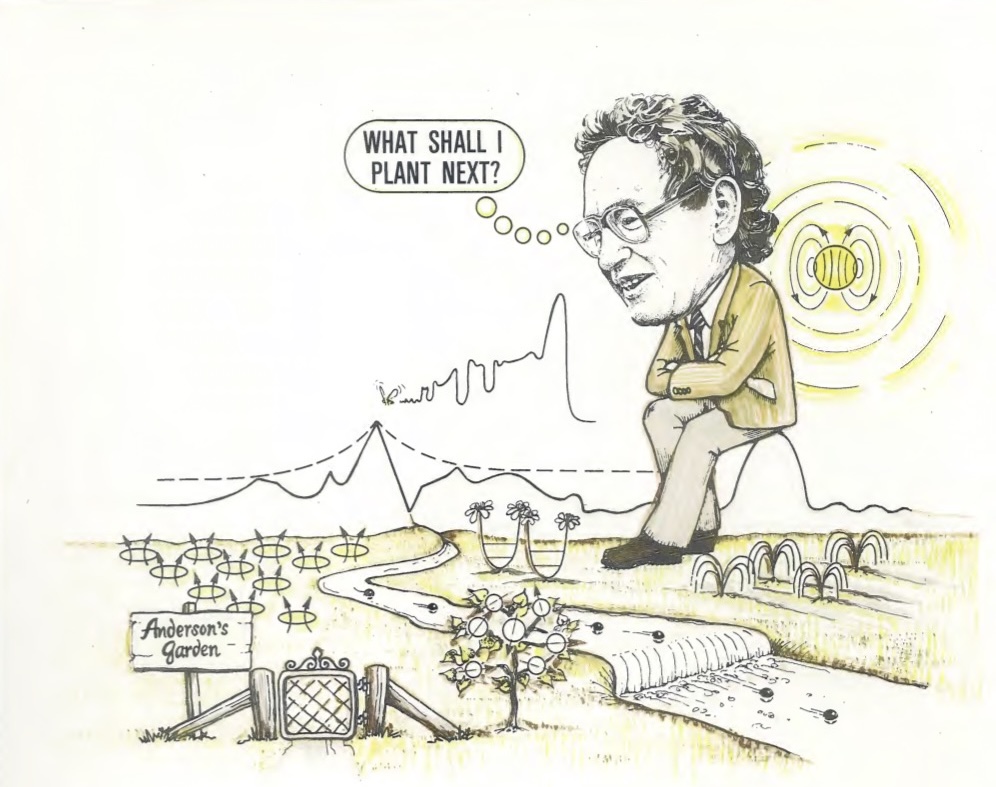}{fig2a}{''Anderson's Garden'', a schematic presented to Phil on the occasion of his 60th birthday, illustrating his garden of scientific accomplishments; Phil was an avid gardener, both
of plants and of ideas. A flux lattice in a superconductor is on the left, on the right is a lattice of vortices in a superfluid.  The mountain range in the background is presumably a pulsar ``glitch''.  Phil proudly hung this drawing on the wall of his Princeton office alongside photographs of Joyce and of the Russian physicist Lev Landau.  (Artist unknown) Reprinted from \citep{Chandra15}.}

The Santa Fe Institute (SFI) was started in 1974 to encourage the flow
of ideas between traditionally disparate fields. Phil was one of
the interdisciplinary scientists who founded the SFI; he later became an
external Professor and served as a Vice-Chairman of its Science Board.
Phil and Nobel economist Kenneth Arrow led SFI workshops that brought
together economists with natural scientists.

\section{{\bf Political and Cultural Activities}} Phil grew up amidst much
political discussion, and he was not afraid stand up for his beliefs.  He often recounted how he was one of the few Bell Labs
scientists who refused to answer a questionnaire about his political
views during the McCarthy period; however, he admitted that Joyce was
worried about possible consequences even though she shared his opinions.
Later in life, Phil used the visibility of his Nobel Prize to take
public stands. For example he signed letters protesting the
incarceration of the Russian dissident physicist Yuri Orlov (1978),
endorsing a Comprehensive Test Ban Treaty (1999) and opposing the Iraq
War (2003) \citep{3Zangwill}. Phil was particularly proud of his opposition to SDI,
the Strategic Defensive Initiative, colloquially dubbed ``Star Wars''.
Phil wrote ``{\sl I happened to be in a position to be caught up in the
campaign against {`}Star Wars' very early (summer {'}83) and
wrote, spoke and testified repeatedly, with my finest moment a debate
with Secretary George Schultz in the Princeton Alumni Weekly, reprinted
in Le Monde in 1987}.'' \citep{3Zangwill} (\ref{Anderson16}) In private discussions Phil often nudged
colleagues, particularly young researchers, to develop opinions and to be ready to defend them.

In addition to his scientific adventures, Phil was well read on a
variety of subjects. Andrew Robinson of the Times Higher
Education Supplement commissioned several of these pieces,
noting that Phil had originally planned to do more of this
as the years progressed.
``{\sl My dream of giving up physics for writing, however},''
wrote Phil in 2011 (\ref{1031Anderson}), 
``{\sl never materialized; physics continues to
dominate my professional life.}" Phil himself was a living refutation of the standard notion
that achievement in physics is age-restricted; he wrote ``{\sl in my experience the cliché is wrong, 
wrong, wrong} (\ref{1031Anderson})'' and ``{\sl to my surprise, physics has remained exciting and wholly
absorbing as I age, and while I love to write, that occupation has
always been secondary}"(\ref{1031Anderson}). 

Phil remained active and opinionated about physics well into his ninth
decade. In 2006 a statistical analysis \citep{32Soler} gave Phil its highest
``creative index'', naming him the ``world's most creative physicist;
always competitive, Phil was thrilled to be followed by particle
theorists Steven Weinberg and Edward Witten in this study. Phil was
curious to the very end. He passed away on March 29, 2020. 
According to his daughter Susan, one of the last
books Phil was reading was ``Birth of a Theorem: A Mathematical Adventure''
by Cedric Villani \citep{33Villani}.\\
%\vfill\eject

\part{{\bf 
Selected Scientific Work}}

\section{{\bf Overview}}

Throughout his academic life, Phil's insatiable curiosity led him to ask
questions that often resulted in new fields of research. He wrote
copiously about his impressions of talks, discussions and papers, and
most importantly on his many creative ideas. His Bell Lab notebooks,
currently housed in the Firestone Library of Princeton University,
present a window into his original 
thinking patterns \ref{notebook}. There are
entries under the heading ``impossible experiments'', such as the
``neutrino Mossbauer effect'' displayed in Figure(\ref{fig3}), where Phil
brainstormed about novel kinds of measurements; also there are instances
where he commented ``Great idea flopped'' and then he moved on to other
things (\ref{notebook}). Here we present a flavour
of his key scientific contributions.

\fg{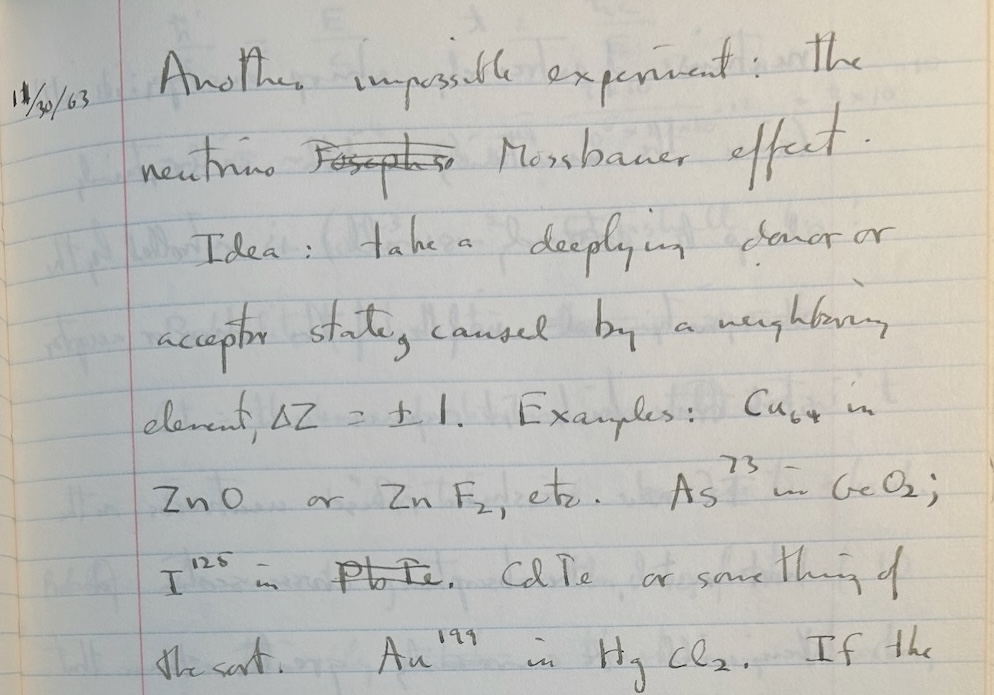}{fig3}{``Another impossible experiment''. Entry from Phil
Anderson\textquotesingle s 1963 notebook on
30\emph{\textsuperscript{th}} November, 1963 discussing the possibility
of a neutrino Mossbauer effect (\ref{notebook}).}
%\includegraphics[width=0.95\textwidth,alt={Extract from Phil Anderson's notebook 11/30/1963}]{media/Notebook_selected_cropped.jpg} ``Another impossible experiment''. Entry from Phil
%Anderson\textquotesingle s 1963 notebook on
%30\emph{\textsuperscript{th}} November, 1963 discussing the possibility
%of a neutrino Mossbauer experiment.

%\section{{\bf Broken Symmetry}}

One of Phil's pioneering contributions to physics was in furthering the
concept of ``broken symmetry'', an idea originated by the Russian physicist, Lev Landau.  Symmetry is a cherished principle in
physics, thus a fluid such as water, has continuous translational and
rotational symmetry because its properties are independent of position
or the direction in which they are observed. However, when water
freezes, its atoms crystallize into a regularly spaced lattice acquiring
discrete rather than translational and rotational symmetries, a feature
responsible for the shape of a snow flake. Thus crystallization
``breaks'' the symmetries of the fluid, but in
return, it develops a beautiful regularity and forms a new state of
matter, a solid.

Phil extensively used the concept of broken symmetry to understand new
phases of matter, including magnetism, superconductivity, superfluidity
and he further extended his insights to particle physics, where his
discovery of the Anderson-Higgs mechanism that gives mass to gauge
bosons, is central to the modern understanding of elementary particle physics 
and the early universe.

\section{{\bf Anderson's Magnetic Life}}

The spin angular momentum of the electron plays a central role in magnetism. The spin an electron is quantized to be either ``up'' or ``down'', denoted
by $\uparrow$ or $\downarrow$. Ferromagnetism, seen in iron, involves the parallel
orientation of electron spins, denoted as ($\uparrow\uparrow\uparrow\uparrow\uparrow$ ), but in an
antiferromagnet, electrons at neighboring atoms alternate between ``up''
and ``down'' directions, denoted by ($\uparrow\downarrow\uparrow\downarrow\uparrow\downarrow$), and the magnetic fields
they produce cancel, leading to no net magnetization. Quantum mechanics causes the spins in an antiferromagnet to fluctuate between configurations,   $(\uparrow\downarrow )\rightleftharpoons(\downarrow\uparrow )$, which could melt the order. 
Why then do materials pick a particular antiferromagnetic arrangement?
Phil showed that the strength of these quantum fluctuations depend on dimensionality (\ref{RN3430}). The fluctuations become so strong in
one dimensional chains that they melt the antiferromagnetic order, but
they become weak enough in higher dimensions to allow
antiferromagnetism to survive, albeit with a 
reduced magnitude (\ref{RN3430}).

In antiferromagnets such as MnO, Phil helped clarify how two magnetic atoms can develop an antiparallel spin alignment, even when they are separated by a non-magnetic atom, such as oxygen, and are thus too far apart to interact via direct exchange. This interaction was termed superexchange. In one of his earliest works at Bell Labs,  Phil showed that when electrons exchange their spins via an intermediate non-magnetic ion, they form a high energy virtual state with two antiparallel electron spins in the 
same orbital (\ref{50AndersonSuperexchange}). These virtual fluctuations stabilize the antiferromagnetic configurations of the electrons on the magnetic atoms. Phil subsequently showed that the virtual charge fluctuations responsible for  antiferromagnetism do not require an intermediate non-magnetic atom (\ref{36Anderson}), thus broadening the meaning of superexchange.  Key to this later work was the adoption of Mott's\citep{mott49} conceptualization of the Coulomb interaction as an onsite repulsion energy, which he labeled ``$U$", and the formulation of his ideas using a second-quantized formalism.

Phil's local magnetic moment model (\ref{37Anderson}), cited prominently by the Nobel Prize committee (\ref{Anderson16}), addresses the question of how a localized electron magnetic moment survives in a metal. Phil's model considered an atomic impurity with a localized electron orbital, such as a transition metal atom with a d-orbital immersed in a metal. Two key  features of the model are the hybridization between the d-orbital and the conduction sea, allowing electrons to tunnel in and out of the d-orbital, and the introduction of the onsite $U$ from his earlier work.  He demonstrated that in the magnetic state, a single electron spin occupies the localized orbital, while additional electrons are blocked from occupying the orbital by the on-site Coulomb repulsion $U$.

\section{{\bf The Kondo Problem}}

Phil's 1961 paper on moment formation left open the question of how a
magnetic moment in a metal  behaves at low temperatures, when quantum fluctuations become important.   This innocent
question blossomed into a major
theoretical challenge known as the Kondo problem.
Most electrical resistance in metals is caused by the scattering of
electrons from thermal vibrations of atoms. Normally, as these vibrations subside upon cooling, the resistivity decreases. However, in the 1930s, experiments found that the resistivity of copper, gold and silver reaches a minimum, and then
starts to rise upon further cooling. In 1964 Myriam
Sarachik at Bell Labs showed experimentally that the mysterious ``resistance minimum'' \citep{38Sarachik} is caused by electrons scattering from magnetic iron impurities.

Around this time, the Japanese physicist Jun Kondo used perturbation theory to study the scattering of electrons off iron
moments. He used a model of spin exchange between conduction electrons and a localized d-electron, a model that is contained in Anderson's local moment model. Kondo showed that the predicted resistance has a minimum \citep{39Kondo}, but there was a problem: below a certain temperature, the ``Kondo temperature'', he found that magnetic scattering becomes so strong that his perturbation theory was no longer valid. What happens at lower temperatures? Nobody knew. 
This question became known as the ``Kondo problem'', and it became a cause célèbre.

The solution to the Kondo problem had to await the development of a new tool  called the ``renormalization group'' that had entered particle and statistical physics in the 1960s and 1970s. Phil's team, consisting of his Cambridge
student, Gideon Yuval and Bell-Labs colleague Don Hamman, successfully applied renormalization to the 
Kondo problem (\ref{40Anderson}, \ref{70AndersonPoorManScaling}) by mapping the sequence of spin flips in time
onto a classical 1D Ising model with long-range interactions.  This work
convincingly demonstrated that at low
temperatures, the local moment forms a singlet with effectively a single conduction electron spin and becomes nonmagnetic, i.e., the conduction electrons completely screen the magnetic
moments. 

The influence of these new ideas ran far and wide, 
%in the
%condensed matter theory community
establishing the importance of
renormalization methods in quantum systems, opening a new path forward for
theoretical work. It influenced Kenneth Wilson \citep{45Wilson} to develop
his numerical renormalization approach, the basis of modern
density-matrix renormalization group approaches to quantum problems
\citep{46White}. The Anderson-Yuval-Hamann approach was later adapted by Kosterlitz
and Thouless \citep{47Kosterlitz} to understand the interaction between vortices in a
two dimensional superfluid, leading to the famous
Berezinskii-Kosterlitz-Thouless phase transition, for which Kosterlitz
and Thouless were awarded the Nobel prize in 2016.

\section{{\bf Disorder-Induced Physics}}

Before Phil came along, most of the work in solid state physics had
centered around ordered, crystalline materials. Phil pioneered the idea
that disorder leads to new physics and should not be thought of as a
merely a perturbation or an annoyance. As we describe below, this can be
seen in his seminal works on localization, local magnetic moments (see
above), glasses and spin glasses.

\subsection{{\bf Localization}}

As far as electrical conduction is concerned, there are two basic types
of materials: insulators and metals. In metals electrons flow in
response to an applied electric field (or voltage), giving rise to an
electric current, but in insulators electrons are not able to flow and
conduct electricity in response to an applied voltage.

By 1958 there was a good understanding of how electrons conduct in
crystalline metals, where the atoms that are arranged in an orderly
fashion. While it was known that disorder in a crystal, e.g., impurities or
atoms displaced from their lattice sites, would lead to increased scattering of moving electrons and increased electrical
resistance, it was generally believed that the electrons would continue to diffuse through the lattice, which would thus remain metallic. However, in a landmark 1958 paper,  entitled, ``Absence of Diffusion in
Certain Random Lattices'' (\ref{48Anderson}),
Phil showed that in the presence of enough disorder, electrons become localized, or confined, to a particular region
of the material so that they no longer conduct, resulting in an
insulator.

Anderson's work defined an entirely new problem, and both his approach
and conclusion are testimonies to his originality and creative insight.
His 1958 localization paper \ref{48Anderson}, was cited as one of the reasons for awarding him the Nobel Prize. Yet, this work was largely unappreciated
(even by the author) and it would be many years before Phil returned to
the problem. 
%(More details are in the supplement.)

In 1979 Phil revisited to the problem of localization. In a landmark
``Gang of Four'' paper (\ref{19Abrahams}), the authors considered disordered metals
in one, two and three dimensions. They used a scaling
theory to show that one and two-dimensional systems would become insulators
for infinitesimal amounts of disorder. This work has had a wide influence in science and
is one of the most cited papers in Physical Review Letters (\ref{19Abrahams}).

The original ideas of Anderson localization were for 
disordered systems in which interactions between electrons are ignored (\ref{48Anderson}). Phil presented qualitative
arguments that such localization should survive weak short-range
interactions (\ref{49Fleishman}). The subsequent observation of metal-insulator
transitions in several dilute two-dimensional electron and hole systems
\citep{50Kravchenko} was thus a surprise to the community, and the study of such
disordered, interacting phases (in thermal equilibrium) remains active
to this day \citep{51Lee,52Giamarchi,53Giamarchi}.

Localization is being actively investigated in the context of (super)
cold atoms where the presence of a disordered potential and interactions
can be carefully controlled. Localization has been experimentally
observed in cold atom systems both in the absence of interactions \citep{54Billy,55Roati} and in the presence of interactions \citep{56Errico}. There is also interest in
so-called many-body localization where a system of interacting electrons
in a disordered potential is prepared in thermal equilibrium with a
thermal bath, and then isolated from the bath and allowed to evolve
quantum mechanically to see if the conductivity is zero or not \citep{57Basko,58Abanin,59Gornyi}.

\subsection{{\bf Two Level Systems}}

In 1971 Zeller and Pohl \citep{60Zeller} found that at temperatures below 1 K,
insulating glassy (non-crystalline) materials, regardless of their
chemical composition, exhibit a specific heat that is linear in
temperature T and a thermal conductivity that goes as
T\textsuperscript{2}. (Specific heat is indicative of the number of
degrees of freedom that can be excited when the system absorbs heat.
Thermal conductivity is a measure of how well the system conducts heat.)
In seminal papers, Anderson, Halperin and Varma (\ref{61Anderson}), and independently,
W. A. Phillips \citep{62Phillips}, proposed that the thermal behavior in glasses at
low temperatures could be explained by so-called two level systems
(TLS). The basic idea is that an atom or group of atoms can sit more or
less equally well in either of two configurations, i.e., in either of
two minima of a double well potential. At low temperatures, there is not
enough thermal energy to hop over the barrier between the two wells, but
quantum mechanically, the atom or group of atoms is able to go through
the barrier, i.e., tunnel back and forth between the two configurations
represented by the two minima. The system has two energy levels. In the
TLS model, the energy difference between these two energy levels have a
uniform or flat distribution up to a few tens of degrees, resulting in a
specific heat that is linear in temperature. Furthermore, the phonons
carrying the heat scatter from the TLS, resulting in a thermal
conductivity that is quadratic in temperature. Although the microscopic
nature of the TLS remains a mystery in most cases, the model has been
enormously successful in explaining a variety of both thermal and
ultrasonic measurements. Indeed, by drawing an analogy between TLS and (nuclear) spins, many of the well-known nuclear magnetic resonance
techniques (NMR) have analogs when the magnetic field is replaced by a
strain field (or an electric field if the TLS have electric dipole
moments), e.g., phonon echoes in glasses are analogous to spin echoes.
Although Phil did not continue to work on TLS, the model and its
implications live on in the present day. Indeed, currently, TLS are a
focus of intense interest in the effort to build a quantum computer from
superconducting and semiconducting qubits because TLS with electric
dipole moments are an important source of noise and decoherence \citep{63Martinis},
and must be dealt with if quantum computers are to become a reality.

\subsection{{\bf Spin Glasses}}

In 1975, Phil and Sam Edwards proposed the model of spin glasses (\ref{117Edwards}). They were inspired by experiments in which
magnetic impurities, such as manganese atoms, were doped into a
non-magnetic host such as copper. As these systems were cooled, they
exhibited a transition into disordered magnetic states.

In ordered magnetic systems such as a ferromagnet or an antiferromagnet, the interactions between pairs of magnetic moments, or spins, are
identical. However, in a spin glass, the interactions are random,
leading to frozen, disordered configurations of spins at low
temperatures. Phil and his collaborators introduced a number of
important theoretical techniques to deal with the challenges that
frozen-in randomness, known as quenched disorder, posed. These include
the so-called `replica trick' (\ref{117Edwards}) and the 
Thouless-Anderson-Palmer (TAP) equations (\ref{77AndersonTAP}). 
Many years later two researchers who worked
in the early stages of their careers with Phil (B.G. Kotliar as a
graduate student and A. Georges as a post-doctoral fellow) built on
these ideas to develop Dynamical Mean Field Theory \citep{64Georges}, a framework for incorporating strong correlations in electronic properties that is
widely used now to study strongly interacting quantum materials.

The impact of the concepts developed for spin glasses extend far beyond
disordered magnets. One of the key concepts was frustration in which a
spin cannot find an orientation that satisfies the demands of all the neighboring spins. For example, one neighbor wants a spin to point up
while another neighbor wants it to point down. Another key concept, also
recognized in structural glasses \citep{Goldstein1969}, is the picture of an
energy landscape where valleys, or minima, represent metastable states.
Simulated annealing was developed as a way to find these local minima.
When John Hopfield was developing his model of neural networks, Phil
encouraged him to consider spin glasses as an analog system where the
spins represent neurons and the connections between neurons were
represented by the random interactions between spins. Other applications
include combinatorial optimization, content-addressable memories, and
most recently, machine learning.

%NP Completeness needs to be added?

\section{{\bf Superconductivity: a Scientific Love Affair}}

Kamerlingh Onnes discovered superconductivity in 1911 when he cooled
mercury and found that its electrical resistance suddenly vanished at a
transition temperature of T$_c$ = \mbox{4.2 K}. It was not until 1957 that John
Bardeen, Leon Cooper and Robert Schrieffer published their
microscopic theory of superconductivity. In the ``BCS theory'', the superconducting wavefunction is written in terms of ``Cooper pairs,'' with each pair consisting of two electrons with opposite spin and opposite momenta. On the day in
spring of 1957 when Phil first heard about BCS at a talk at Princeton by
David Pines, it was ``love at first sight'' (\ref{65Anderson}). On the way back to
Bell Labs, he had a eureka moment that linked BCS theory to magnetism.
Using a ``pseudo-spin'' reformulation of BCS, Phil was able to map the
problem of superconductivity onto magnetism. This new analogy enabled
him to confirm that BCS was gauge invariant and establish that the
Coulomb repulsion between electrons in a superconductor is screened just
as it is in a metal (\ref{66Anderson}).

Anderson's reformulation of BCS had a wide influence. By linking
superconductivity and magnetism, the community took a cautious step
closer to regarding the phase of the superconducting wavefunction as a
palpable, detectable variable (with the caveats of gauge fixing). This
emerging perspective would culminate in the discovery of the Anderson-Higgs mechanism (67) and the Josephson effect(\ref{13Anderson},\ref{68Anderson}).

\subsection{{\bf Josephson Effect and Broken Symmetry}}

As we described earlier, Brian Josephson was a young graduate student in Phil's course when he came up with his Nobel-Prize winning ideas(\ref{68Anderson})\citep{Josephson1962}. Phil was probably among the first to appreciate Josephson's ideas, 
and they had a profound influence on his evolving understanding of superconductivity and broken symmetry\citep{12Anderson}(\ref{13Anderson}).

The Josephson effect arises in a tunnel junction between two
superconductors. A tunnel junction involves a nanometer-thin insulating layer, usually formed from an oxide, sandwiched between two superconductors. Each superconductor has a coherent superconducting
wavefunction consisting of superconducting electrons. Think of the
wavefunction as a wave with an amplitude and a phase. If, as is usually
the case, the two superconductors have different phases, then current
will flow across the tunnel junction without any applied voltage drop,
i.e., without any batteries! The current depends nonlinearly on the
phase difference \(\delta\), and is proportional to \(\sin\delta\).
Josephson also predicted that if a voltage is applied across the
junction, then the phase would vary in time.

When Phil returned to Bell Labs in August of 1962, he began to discuss
with his colleague, John Rowell, the possibility of experimentally
confirming Josephson's effect, initiating a very close collaboration
between experimentalist and theorist. In January 1963 when Rowell first
tried the experiment, no supercurrent was observed. Rowell recalls that
Phil realized overnight that \citep{69Rowell}:

\emph{``\ldots I had to make lower resistance junctions\ldots{} So I
tried a tin-lead junction first, this was January 21\textsuperscript{st}
1963 that I first saw a convincing Josephson current''.}

\noindent The new experiment displayed currents with tremendous sensitivity to
magnetic fields.

Their resulting paper had Phil as the first author \citep{70Rowell,71Rowell}. Anderson
and Rowell went on to propose and patent the use of superconducting
quantum interference effects as a method for detecting minute magnetic
fields.
%\sout{The Josephson effect has many important applications. For example, it is
%a way to accurately measure the ratio h/e where h is Planck's constant
%and e is the magnitude of the charge on an electron.} 
The modern realization of this idea
is the (DC) superconducting quantum
interference device (SQUID), which consists of a loop of
superconducting wire with two Josephson junctions.
%{This is the analog of the two slit experiment.} 
Just as a wave passing through two slits in a wall produces constructive and destructive interference patterns, 
the superconducting wavefunction going through the two
junctions leads to interference that is sensitive to the magnetic flux through the loop, creating an extremely sensitive magnetic field detector. Today, SQUIDs are  used to detect the magnetic fields produced by
electric currents in the brain and are the leading
candidate for quantum bits, or qubits, which are the basic
element of a quantum computer \citep{72Levi}.

The Josephson effect showed that a superconductor prefers a smooth
wavefunction with the same phase everywhere: if
there are variations in the phase, then current flows to eliminate
the phase difference, even with an insulating tunnel junction in the way \ref{13Anderson}. 
This  puzzled the community, and luminaries, such as John Bardeen \citep{BardeenJosephson} initially opposed it, because it went against a belief that  the phase of the quantum
mechanical wavefunction is unmeasurable and therefore without physical consequences.
{%Piers: The idea of phase rigidity goes back to London's 1937 paper, and then onwards to Landau and Ginzburg. It certainly predates Phil's work. Phil is not uniquely responsible for the idea of broken symmetry.  Somewhere in our paper we need to explain,
The new insight was that the energy can be a function of the \underline{gauge invariant} difference of phases \begin{equation}
    E \sim \biggl (\vec\nabla  \phi - \frac{2 e}{\hbar }\vec A\biggr)^2 \rightleftharpoons  -\cos \left(\phi_2-\phi_1 - \frac{2 e}{\hbar} \int_1^2 {\vec A} \cdot { d\vec \ell}\right),
\end{equation} 
where $\phi$ is the phase of the superconductor while $\vec A$ is the electromagnetic vector potential; the left-hand equation is the gauge invariant energy to twist the phase of a superconductor in the bulk that had been introduced by Ginzburg and Landau\citep{ginzburglandau},  whereas the right-hand equation, is Josephson's expression for the gauge invariant energy for two superconductors with a phase difference between their phases $\phi_2-\phi_1$,  separated by an insulating barrier. The important point was that the combination of the phases and the electromagnetic field $\vec A$ \underline{is} gauge invariant.   
Josephson's results had a profound influence on Phil: % It was certainly Josephson who saw this first, and it must have been this that they talked about when they set $\pi = i=e$ in the early sixties at Cambridge, but it was surely 
he internalized them and realized their broader implications,  not only for superconductors but also for broken gauge symmetry in particle physics\citep{12Anderson}. As we will discuss shortly, it was this new perspective that allowed him in the autumn of 1962,  to  propose the Anderson-Higgs effect \ref{98AndersonHiggs}. 
%This came to be known as phase rigidity and led Phil to realize the general principle of broken symmetry \citep{12Anderson}. 

\subsection{{\bf A Mile of Dirty Lead Wire}}

Although superconductors involve a quantum mechanical phase coherence of
electrons, they are unexpectedly robust against disorder. Phil often
quoted Hendrik Casimir's remark, that ``\emph{The remarkable thing is
that electrons somehow manage to maintain a kind of order, whatever it
may be, over a mile of dirty lead wire''}(\ref{74Anderson}). He recognized that the
robustness of superconductivity against disorder results from time-reversal
symmetry, the invariance of physics under the reversal of a trajectory.
In his famous ``dirty superconductor theorem'', published in 1959 (\ref{75Anderson},
he demonstrated that Cooper pairing survives without a reduction of T$_c$,
so long as the disorder preserves time-reversal symmetry.

Another puzzle was how electron pairing in BCS theory could survive
Coulomb repulsion. Lattice vibrations, or ``phonons'' provide the
pairing glue of BCS theory. BCS theory tacitly assumed that the
electron-phonon attraction was big enough to overwhelm the Coulomb
interaction, but did not provide a mechanism. In 1960, Pierre Morel, who
was a French graduate student, and Anderson recognized that the key to
the puzzle lies in the very different time-scales associated with the
two interactions. Electrostatic repulsion is instantaneous, but the
electron phonon interaction is ``retarded'', because ions in a crystal
respond much more slowly than electrons, so the deformation of the
lattice created by a passing electron lingers long after it has passed
by, which allows it to attract electrons long after the Coulomb
repulsion has died away. Anderson and Morel showed that the retarded
part of the Coulomb repulsion is much reduced, and it is this reduction
that makes superconductivity possible (\ref{76Morel}) \citep{77Tolmachev}.

\subsection{{\bf Flux Creep}}

Superconductors expel magnetic fields from their interiors if the fields
are not too large. This is known as the Meissner effect. If the external
field is too large, then superconductivity is destroyed in type I
superconductors. However, as the field increases, type II
superconductors admit larger fields in the form of magnetic vortices or
magnetic flux lines. These magnetic filaments can be thought of as tiny
tornadoes where there is electric current swirling instead of
wind. The core of the vortex is basically a normal metal wire with
normal (non-superconducting) electrons. As the field increases, more
vortices are admitted and eventually the system becomes a normal metal
when the vortices overlap. When electric current flows, these vortices
are pushed sideways and if they move, the normal electrons in their
cores produce electrical resistance. Pinning a vortex prevents it from moving.
Phil pointed out that it is not
necessary to pin each and every vortex because if the vortices are close
enough to each other, they will move in ``flux bundles,'' a concept
introduced by Anderson (\ref{78Anderson}). Anderson pointed out that the pinning of
flux bundles prevents them from moving easily; this leads to ``flux
creep.'' Type II superconductors are used commercially to produce high
magnetic fields because they can carry large currents as long as the
flux lines are effectively pinned. MRI machines used in the clinic are a
prime example. 
%(More details can be found in the supplement.)

\subsection{{\bf Resonating} {\bf Valence} {\bf Bonds}}

The chemist Linus Pauling had introduced the concept of a resonating
valence bond in structures such as benzene molecules, and had actually
suggested this as a basis for understanding the metallic state \citep{79Pauling}.
Phil, working with postdoc Patrick Fazekas, now applied this concept,
arguing that in certain frustrated two dimensional magnetic insulators,
such as a triangular lattice antiferromagnet, the ground-state would
consist of a fluctuating liquid of spin singlets formed between
neighboring spins. They called it a ``resonating valence bond'' or RVB
state (\ref{80Fazekas}).

This idea received very little attention until 1986, when the discovery
of high temperature superconductivity by Bednorz and Muller changed
everything \citep{81Bednorz}. Bednorz and Muller found that doping strontium into
insulating lanthanum cuprate yields a high temperature superconductor.
Insulating lanthanum cuprate is a type of Mott insulator, in which the
strong Coulomb repulsion between electrons produces an insulator; that
such a system should become a high temperature (high T$_c$) superconductor
upon doping was truly mind-boggling.

Upon hearing of the new discoveries, Phil had immediately started to
wonder whether insulating lanthanum cuprate might be an RVB spin liquid
with pre-paired electrons. No charge could move in the spin liquid, but
on doping with strontium, electrons removed from the liquid would create
mobile vacancies, or holes, transforming the pre-paired state into a
superconductor. Unlike conventional superconductors with Cooper pairs,
Phil introduced the new idea that a pre-paired fluid of electrons was
already present in the insulator. The resulting paper in Science is one
of his most highly cited papers(\ref{82AndersonRVB}.

Surprisingly, Anderson abandoned his idea of RVB superconductivity in
the early 1990s, only to change his mind again in 2004 as it became
increasingly clear that the idea may indeed contain a strong element of
the as yet, undiscovered final theory of high-T$_c$ (\ref{83AndersonRVBvanilla}. Indeed, early
work by one of Phil's former students, Gabriel Kotliar \citep{84KotliarRVB,85KotliarLiu},
demonstrated that RVB theory predicts the d-wave pairing
symmetry observed in high-T$_c$ cuprate superconductors. More generally RVB
theory seeded a new genre of
research: it inspired a search for spin liquids \citep{86Balents,87Knolle} and sparked a
new interest into emergent gauge fields in quantum materials, a
prominent area of current research \citep{88Sachdev}.

\section{{\bf Superfluid Helium-3}}

Sometime around 1960, physicists realized that BCS theory could be generalized to Cooper pairs of fermions with a finite angular momentum, and this
led to the prediction that at low temperatures, helium-3 would become a superfluid, a neutral fluid
with zero viscosity. Like a molecule, Cooper pairs can have integer orbital angular momentum (s-states ($\ell=0$), p-states ($\ell=1$), and d-states ($\ell=2$). 

In a paper with Brueckner and Soda, Anderson and Morel predicted that helium-3 would develop a d-wave or $\ell=2$ superfluid (\ref{89Brueckner}). In a second paper, Anderson and Morel extended the theory further to the p-wave case in which the spins of the fermion pairs align to form a triplet. This led them to discover a magnetic superfluid ground-state (\ref{90AndersonHe3}).

Some 13 years later, Lee, Osheroff, and Richardson discovered superfluid helium-3 at 1 mK; surprisingly, there are two superfluid phases
that they called the ``A'' and ``B'' phases \citep{91Osheroff}. It turns out that the helium-3 atoms do not
pair in a d-state but in a p-state. Later analysis showed that the A-phase corresponds to the Anderson-Morel phase
that they proposed in 1961. The B-phase corresponds a state
proposed in 1963 by Balian and Werthamer \citep{92LegettHe3}.

\section{{\bf The Anderson-Higgs Effect}}

In the early 1960s particle physicists were puzzled by the short-range of
the nuclear forces that hold protons and neutrons together in a nucleus. Yukawa
and others had reasoned that these forces are mediated by the exchange
of massive particles. On the other hand, theoretical considerations
suggested that these exchange forces would be carried by nuclear analogs
of the photon, called gauge bosons, and that this would inevitably
result in massless particles with long-range nuclear forces. Since the
forces of particle physics are short-ranged: this posed a conundrum. In
a remarkable set of insights based on superconductivity, Anderson proposed a way out \citep{96WittenPWA,97CloseHiggs} \ref{98AndersonHiggs}.

 Anderson realized that photons acquired a mass in superconductors and this made the electromagnetic force inside a superconductor short-ranged (\ref{99AndersonRPA}. A manifestation of this is the Meissner effect in which external magnetic fields decay exponentially at the surface of a superconductor, which can be viewed as a massive photon. 

From his interactions with Brian Josephson as well as with Bell Lab visitors, Robert Brout
and Yoichi Nambu, Anderson 
proposed a mechanism in which the short-range nuclear forces are mediated by subatomic gauge bosons that acquire mass in a kind
of cosmic superconductor \citep{12Anderson}, known today as the Higgs field after the
British physicist Peter Higgs. Anderson's seminal 1962 paper ``Plasmons, gauge invariance and mass'' (67) was
prominently cited in a 1964 paper by Peter Higgs which predicted the Higgs boson (\ref{98AndersonHiggs}.

Two years later, a series of particle physicists, Peter Higgs, Robert
Brout and Francois Englert, Tom Kibble and others, generalized the
Anderson's mechanism to a relativistic theory applicable to particle
physics \citep{96WittenPWA}. Both Higgs and Kibble mention Anderson's work in their
seminal papers. Peter Higgs and Francois Englert received
the Nobel prize for this work in 2018. In a nod to Phil's achievements,
%condensed matter 
physicists now refer to the mechanism by which gauge
bosons acquire mass as the Anderson-Higgs effect.

\section{{\bf Astrophysics: Neutron Stars}}

When stars undergo a supernova explosion, protons and electrons are
crushed together by the recoil, leaving behind a small dense core of
neutrons known as a neutron star. According to
M. Ali Alpar, one of Phil's former graduate students, 
``{\sl Phil's interest in neutron stars
started in 1967 with the discovery of pulsars by Jocelyn Bell, a student
of Anthony Hewish in the Cambridge Radio Astronomy group,}''. Jocelyn Bell had detected intermittent radio signals that occurred
with amazing regularity. These pulses turned out to come from a rapidly
rotating neutron star, termed a ``pulsar'', that emits a beam of
electromagnetic radiation like a lighthouse. We detect this radiation
when the beam sweeps past the earth. The discovery of pulsars was the
first observation of neutron stars, an extreme form of matter that
naturally attracted the attention of several solid-state theorists \citep{94BaymBCS}.

Because neutrons, like electrons, are fermions, BCS theory could be
applied to a superfluid of neutrons in neutron stars, and this
fascinated Phil. He was particularly interested in observed ``pulse
glitches'', temporal irregularities in the emitted radiation. In a
series of papers Phil and his collaborators made strong analogies
between superfluid vortex dynamics in neutron stars and flux dynamics in
type II superconductors. More specifically they suggested that the
observed pulse glitches resulted from the sudden release of neutron
vortex lines that were pinned by nuclei in the neutron star's inner crust \citep{93Alpar}. Though Phil felt that these ideas ``{\sl ended up being right and
probably even being proved out}'' \citep{12Anderson}, the 
``{\sl apathetic response to our work in the astrophysics community}''(\ref{95AndersonNeutron})
may have led him to leave this area.

\section{{\bf More Is Different: Emergence}}

In the 1960s a prevailing philosophy in physics was
reductionism -- the idea that if you reduce everything to its
smallest components, you could then reconstruct and understand
everything. Phil disagreed strongly; in 1969 he gave a lecture at the
University of California, San Diego (\ref{23Anderson} introducing a radically new
perspective which he later published in a landmark Science article, ``More is Different''(\ref{23Anderson}. Phil
argued that reductionism fails to take into account complexity, broken symmetry and length
scales. Instead, he argues that science is hierarchical and that ``At each stage, entirely new laws, concepts and generalizations are necessary, requiring inspiration and creativity to just as great a
degree as in the previous one''. ``More is Different'' ends with a key
assertion ``Surely there are more levels of organization between human
ethology than there are between DNA and quantum electrodynamics, and
each level can require a whole new conceptual structure.''

Curiously, although the word ``emergence'' does not appear in ``More is
Different,'' Phil's article contains several examples of how complexity gives rise to emergent properties, highlighting the importance of
symmetry and the ways it can be broken in the thermodynamic limit. For example, we may know everything about a helium atom but we would be hard pressed to predict superfluid helium from this knowledge.
Anderson dared to go far beyond examples from physics to touch on the social and biological sciences, where he
suggests that DNA might be an example of an ``Information bearing
crystallinity'' and discusses the possible connection of life to the
emergence of states of matter that develop ``temporal regularity''. Over the years ``More is Different'' has become the  clarion call for
the fundamental importance of complexity and emergence, exerting
an enormous impact on research far beyond physics, including biology,
complexity science, non-linear dynamics, economics, philosophy and the social sciences \citep{100StumpfMore,101StrogatzMore,102DosiMore}.

%\section{{\bf Other Contributions}}
%Due to space limitations, we were not able to %cover other contributions that Phil made such as %negative U centers, NP completeness, etc.

\part{{Finale}}

In concluding this Biographical Memoir, we should like to quote from the citation of
Philip Warren Anderson's 1991 honorary doctorate from Rutgers University; it was written by one of
his longtime friends and collaborators,  Elihu Abrahams:

{\sl ``Phil Anderson, your remarkable insights into the
quantum-mechanical foundations of the behavior of matter have transformed
the main lines of thought and practice in modern solid state and condensed matter physics...Your
work shows that the most original and beautiful theoretical concepts
arise from a deep understanding of the phenomena revealed in experimental
laboratories. As the mentor of many junior colleagues who have themselves made outstanding contributions, you have inspired a generation of physicists who champion this dynamic relationship of theory and experiment. The profound discoveries you have made have been generated by
your unique creative personality and confirm the place of science among
the great achievements of the human spirit.''}

\section*{Acknowledgements}

{The authors are very grateful to Susan Anderson for sharing photographs and reminiscences of her father. We would also like to thank Firestone Library at Princeton University for providing access to Phil's archival notebooks.  Many of Phil's colleagues, students, postdocs and friends have contributed to this memoir, and we deeply appreciate your recollections and photos of dear Phil. Finally, we should like to thank the hospitality of the Aspen Center for Physics, which is supported by National Science Foundation grant PHY-2210452, where the final stages of writing this Memoir took place.  }

\section*{Awards and Recognition}

\begin{itemize}\itemsep=0pt
\item[]American Academy of Arts and Sciences, 1963.
\item[]Oliver E. Buckley Prize in Condensed Matter Physics, 1964. 
\item[]Fellow, National Academy of Sciences 1967.
\item[]Dannie Heineman Prize, 1975.
\item[]Nobel Prize in Physics, 1977.
\item[]Honorary Fellow of Jesus College, Cambridge, 1978.
\item[]Honorary Doctor of Science, University of Illinois, Urbana-Champaign, 1978.
\item[]Foreign Member of the Royal Society (ForMemRS), 1980. 
\item[]National Medal of Science, 1982.
\item[]Honorary member, Japan Academy, 1988.
\item[]Member, American Philosophical Society, 1991.
\item[]Honorary Doctor of Science, Rutgers, the State University of New Jersey, 1991.
\item[]Honorary degree, Ecole Normale Superieure, Paris, 2002. 
\item[]Honorary degree, Tokyo University, 2002. 
\item[]Honorary degree, Tsinghua University, 2007.
\end{itemize}

\figwidth=0.9\textwidth
\section*{Author Profiles}
\fg{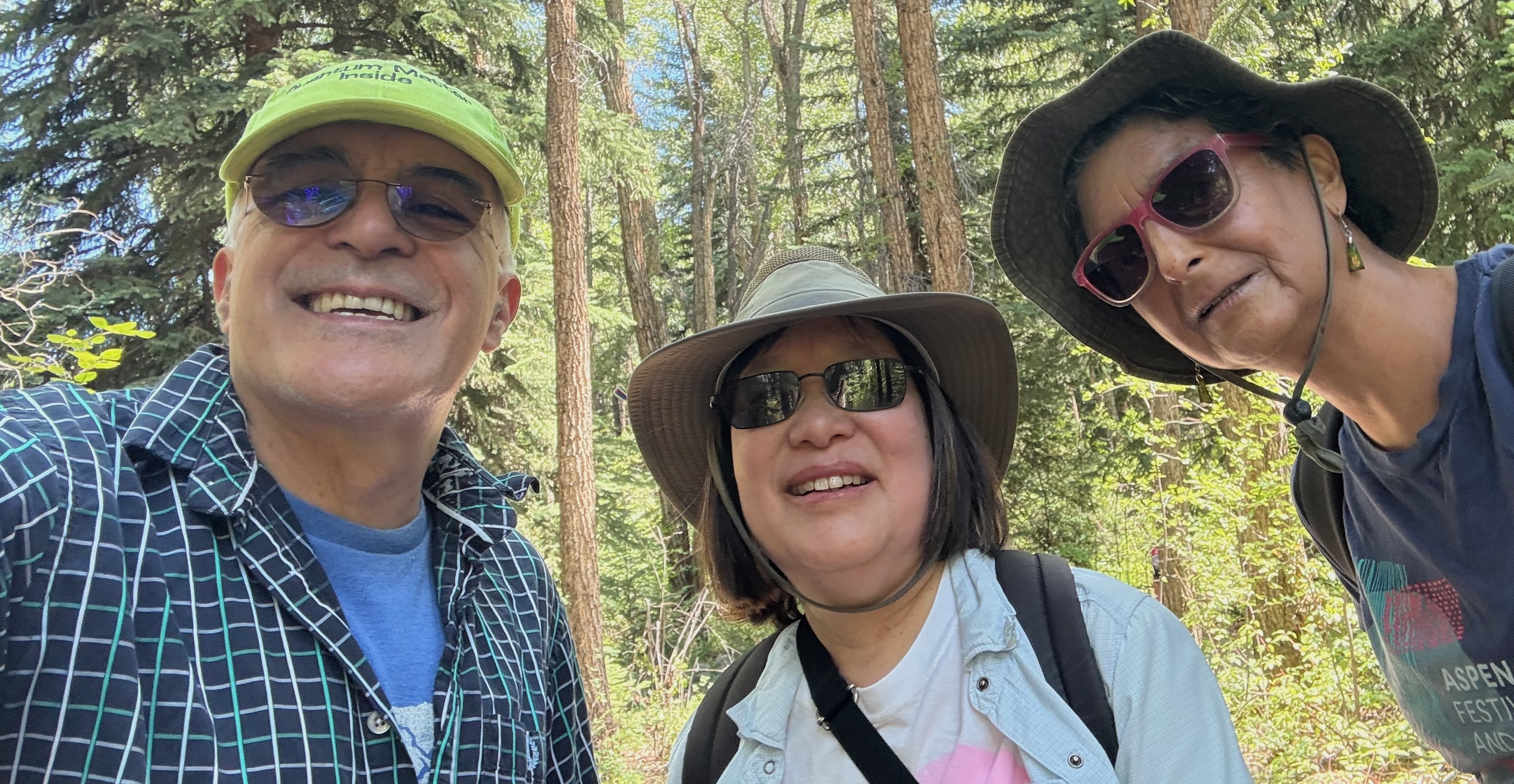}{Fig5}{The three authors in Aspen, Colorado: (from left to right) Piers Coleman, Clare Yu and Premala Chandra
who were all graduate students working with Phil Anderson
in the 1980s.}

\hskip 0.25in{\bf Premala ``Premi'' Chandra} is an materials-inspired theorist who is a Professor in the Department of Physics and Astronomy at Rutgers University (US). During her doctoral studies, Premi was welcomed as a visiting student in Phil's condensed matter group; there she became interested in frustrated magnetism, a research area that continues to fascinate her to this day. Premi maintained contact, both scientifically and socially, with Phil from that time onward. Premi is a Fellow of the Institute of Physics (2004) and of the American Physical Society (2013).

{\bf Piers Coleman} is a theoretical physicist, educated at Cambridge and Princeton. Under Phil Anderson, he developed the slave‑boson method in
the 1980s—a powerful technique for modeling heavy‑fermion materials and
unconventional superconductivity. As a Distinguished Professor at
Rutgers and the University of London (Royal Holloway), he has made
landmark contributions across topics such as quantum criticality,
topological Kondo insulators, predicting the gapless surface states of SmB$_6$, and frustrated magnetism. Coleman is director of the International Institute for Complex
Adaptive Matter and the Hubbard Theory Consortium in London.   He is the recipient of a Sloan Fellowship (1988) and a Fellow of the 
American Physical Society (2002). Coleman's ``Introduction to Many‑Body Physics''
(Cambridge Univ. Press, 2015) is a widely read text in the field. 

{\bf Clare Yu} is a Distinguished Professor of Physics and Astronomy at the University of California, Irvine (UCI). A Princeton alumna (AB 1979, PhD 1984), she did both her senior thesis and PhD thesis with Phil Anderson. She credits Phil with teaching her how to think about physics. Her research in theoretical condensed matter physics includes disordered systems and microscopic noise mechanisms that limit coherence in Josephson junction qubits. Her research group also applies statistical physics and simulation techniques to biological questions, including intracellular transport mechanisms and the spatial distribution of immune cells within the tumour microenvironment. She is a fellow of the American Physical Society (2005), the American Academy of Arts and Sciences (2019), and the American Association for the Advancement of Science (2021). 

\section*{References}

%
% To be listed in chronological order...
%

\begin{simplebib}

% (\ref{RN3430},\ref{50AndersonSuperexchange})

\simplebibitem{5Anderson}(1949)
Pressure Broadening in the Microwave and Infra-Red Regions.
\emph{Physical Review} {\bf 76}, 647--661.
(doi: \doi{10.1103/PhysRev.76.647})

\simplebibitem{50AndersonSuperexchange}(1950)
Antiferromagnetism. Theory of Superexchange Interaction,
\emph{Phys. Rev.} {\bf 79}, 350--356.
(doi: \doi{10.1103/PhysRev.79.350})

\simplebibitem{RN3430}(1952)
An Approximate Quantum Theory of the Antiferromagnetic Ground State. \emph{ Phys. Rev.} {\bf 86}, 694–701.
(doi: \doi{10.1103/PhysRev.86.694})

\simplebibitem{48Anderson}(1958)
{Absence of Diffusion in Certain Random Lattices}. \emph{
Physical Review} {\bf 109}, 1492--1505.
(doi: \doi{10.1103/PhysRev.109.1492})

\simplebibitem{66Anderson}(1958)
Coherent Excited States in the Theory of Superconductivity:
Gauge Invariance and the Meissner Effect. \emph{ Physical Review} {\bf  110},
827--835.
(doi: \doi{10.1103/physrev.110.827})

\simplebibitem{99AndersonRPA}(1958)
Random-Phase Approximation in the Theory of
Superconductivity. \emph{ Phys. Rev.} {\bf 112}, 1900–1916.
(doi: \doi{10.1103/PhysRev.112.1900})

\simplebibitem{36Anderson}(1959)
New Approach to the Theory of Superexchange Interactions.
\emph{ Physical Review} {\bf 115}, 2–13.
(doi: \doi{10.1103/PhysRev.115.2})

\simplebibitem{75Anderson}(1959)
Theory of dirty superconductors. \emph{ Journal of Physics
and Chemistry of Solids} {\bf 11}, 26 -- 30.
(doi: \doi{10.1016/0022-3697(59)90036-8})

\simplebibitem{89Brueckner}(1960)
(With K.~A.~Brueckner, T.~Soda, and P.~Morel)
Level Structure of Nuclear Matter and Liquid ${\mathrm{He}}^{3}$. 
\emph{Phys. Rev.} {\bf 118}, 1442–1446.
(doi: \doi{10.1103/PhysRev.118.1442})

\simplebibitem{37Anderson}(1961)
Localized Magnetic States in Metals. \emph{ Physical Review}
{\bf 124}, 41 -- 53.
(doi: \doi{10.1103/physrev.124.41})

\simplebibitem{90AndersonHe3}(1961)
(With P.~Morel)  Generalized Bardeen-Cooper-Schrieffer States and
the Proposed Low-Temperature Phase of Liquid He-3. \emph{Physical Review}, {\bf 123}, 1911--1911.
(doi: \doi{10.1103/PhysRev.123.1911})

\simplebibitem{76Morel}(1962)
(With P.~Morel)  {Calculation of the Superconducting State
Parameters with Retarded Electron-Phonon Interaction}. \emph{ Physical Review}
{\bf 125}, 1263 -- 1271.
(doi: \doi{10.1103/physrev.125.1263})

\simplebibitem{78Anderson}(1962)
  Theory of Flux Creep in Hard Superconductors. \emph{ Physical
  Review Letters} {\bf 9}, 309.
(doi: \doi{10.1103/PhysRevLett.9.309})

\simplebibitem{67Anderson}(1963)
  Plasmons, Gauge Invariance, and Mass. \emph{ Phys. Rev}
  {\bf 130}, 439 -- 442.
(doi: \doi{10.1103/physrev.130.439})

\simplebibitem{68Anderson}(1963)
 (With J.~M.~ Rowell)  Probable Observation of the Josephson
  Superconducting Tunneling Effect. \emph{ Physical Review Letters} {\bf 10},
  230--232.
(doi: \doi{10.1103/PhysRevLett.10.230})

\simplebibitem{anderson1972concepts}(1963)
  Concepts in Solids: Lectures on the Theory of Solids. 
W. A. Benjamin, NY. 

\simplebibitem{notebook}(1963)
Scientific Notebooks.
\emph{Special Collections, Princeton Firestone Library, Princeton}, 
{\bf 1946-2015}.

\simplebibitem{41Anderson}(1968)
  The Kondo Effect. I. \emph{ Comments on Solid State Physics}
  {\bf I}, 31--31.

\simplebibitem{42Anderson}(1968)
  The Kondo Effect. II. \emph{ Comments on Solid State Physics}
  {\bf I}, 190--190.

\simplebibitem{13Anderson}(1970)
  How Josephson discovered his effect. \emph{ Physics Today}
  {\bf 23}, 23--29.
(doi: \doi{10.1063/1.3021826})

\simplebibitem{40Anderson}(1970)
 (With G.~Yuval and D.~R.~ Hamann)  Exact Results in the Kondo Problem. II.
  Scaling Theory, Qualitatively Correct Solution, and Some New Results on
  One-Dimensional Classical Statistical Models. \emph{ Phys. Rev. B} {\bf 1},
  4464–4473.
  (doi: \doi{10.1103/PhysRevB.1.4464})

\simplebibitem{70AndersonPoorManScaling}(1970)
A poor man's derivation of scaling laws for the Kondo problem.
\emph{Journal of Physics C: Solid State Physics} {\bf 3}, 2436.
(doi: \doi{10.1088/0022-3719/3/12/008})

\simplebibitem{43Anderson}(1971)
The Kondo Effect III: The Wilderness-Mainly Theoretical.
\emph{ Comments on Solid State Physics} {\bf 3}, 153--153.

\simplebibitem{23Anderson}(1972)
More is Different. \emph{ Science}, {\bf 177}, 393--393.
(doi: \doi{10.1126/science.177.4047.393})

\simplebibitem{61Anderson}(1972)
 (With B.~I.~Halperin and C.~M.~ Varma)  Anomalous low-temperature thermal
  properties of glasses and spin glasses. \emph{ Phil. Mag.} {\bf 25}, 1--9.
(doi: \doi{10.1080/14786437208229210})

\simplebibitem{44Anderson}(1973)
  Kondo Effect IV: Out of the Wilderness. \emph{ Comm. S. St.
  Phys.} {\bf 5}, 73.

\simplebibitem{80Fazekas}(1974)
 (With P.~Fazekas)  On the ground state properties of the anisotropic
  triangular antiferromagnet. \emph{ Philos. Mag.} {\bf 30}, 423--440.
  (doi: \doi{10.1080/14786439808206568})

\simplebibitem{117Edwards}(1975)
 (With S.~Edwards)  Theory of spin glasses. \emph{ Journal of Physics
  F: Metal Physics} {\bf 5}, 965.
(doi: \doi{10.1088/0305-4608/5/5/017})

\simplebibitem{Anderson16}(1977)
The Nobel Prize: Philip W. Anderson Biographical. \emph{Nobel lectures in Physics 1971-1980}, {\bf 5}, 371--376.
World Scientific.
  
\simplebibitem{77AndersonTAP}(1977)
(With D. J. Thouless and R. G. Palmer), Solution of 'Solvable model of a spin glass'.
\emph{The Philosophical Magazine: A Journal of Theoretical Experimental and Applied Physics},
{\bf 35}, {593-601}.
(doi: \doi{10.1080/14786437708235992})

\simplebibitem{19Abrahams}(1979)
(With E.~Abrahams,  D.~C.~ Licciardello and  T.~V.~Ramakrishnan,
Scaling Theory of Localization: Absence of Quantum Diffusion in Two
Dimensions. \emph{Phys. Rev. Lett.} {\bf 42}, 673-676 (1979).
(doi: \doi{10.1103/PhysRevLett.42.673})

\simplebibitem{49Fleishman}(1980)
 (With L.~Fleishman)  Interactions and the Anderson transition. \emph{
  Physical Review B} {\bf 21}, 2366--2377.
(doi: \doi{10.1103/PhysRevB.21.2366})

\simplebibitem{9Anderson}(1979)
Some memories of developments in the theory of magnetism.
\emph{ J. Appl. Phys.} {\bf 50}, 7281--7284.
(doi: \doi{10.1063/1.326937})

\simplebibitem{20Anderson}(1984)
\emph{ Basic Notions of Condensed Matter Physics}.
Benjamin Cummings.

\simplebibitem{82AndersonRVB}(1987)
The Resonating Valence Bond State in La$_2$CuO$_4$ and
  Superconductivity. \emph{ Science} {\bf 235}, 1196.
(doi: \doi{10.1126/science.235.4793.1196})

\simplebibitem{28Anderson} (1992)
  Complexity II: The Santa Fe Institute. \emph{ Physics Today}
  {\bf 45}, 9--11.
(doi: \doi{10.1063/1.2809684})

\simplebibitem{21Anderson}(1994)
\emph{A Career in Theoretical Physics},
World Scientific.
(doi: \doi{10.1142/9789812385123})

\simplebibitem{95AndersonNeutron}(1994)
(With M.~Alpar, D.~ Pines and J.~ Shaham),  The Rheology of Neutron Stars:  Vortex Line Pinning in the Crust Superfluid,
in (\ref{21Anderson}).

\simplebibitem{83AndersonRVBvanilla}(2004)
(With P.~A.~Lee, M.~ Randeria, T.~M.~Rice, N.~Trivedi and F.~C.~Zhang)  The
  physics behind high-temperature superconducting cuprates: the ‘plain
  vanilla’ version of RVB. \emph{ J. Phys. Condens. Matter} {\bf 16}, R755-- R769.
(doi: \doi{10.1088/0953-8984/16/24/r02})

\simplebibitem{65Anderson}(2010)
{BCS: The scientific ``Love of my Life"} in \emph{BCS: 50 years},127-142, 
World Scientific.
(doi: \doi{10.1142/9789814304665\_0008})

\simplebibitem{22Anderson}(2011)
 \emph{More and different notes from a thoughtful curmudgeon},
World Scientific.
(doi: \doi{10.1142/8141})

\simplebibitem{6Anderson}(2011)
Scientific and Personal Reminiscences of Ryogo Kubo, 
 p 62--67 in (\ref{22Anderson}). 

\simplebibitem{1031Anderson}(2011)
Physics and Bell Labs 1949-1984: Young Turks \& Younger Turks, 
 p1 in (\ref{22Anderson}). 

\simplebibitem{74Anderson}(2011)
 A mile of dirty lead wire: A fable for the Scientifically Literate, 
 p50 in  (\ref{22Anderson}). 

\simplebibitem{98AndersonHiggs}(2015)
Higgs, Anderson and all that. \emph{ Nature Physics}
{\bf 11}, 93--93.
(doi: \doi{10.1038/nphys3247})

\end{simplebib}

\renewcommand{\refname}{References to Other Authors}

% Piers Coleman: Using modified RSBMbibX.bst file, changed to allow doi entries, including for book and incollection
\bibliographystyle{RSBMbibX}
\bibliography{PWA}

\end{document}